 \documentclass[12pt,preprint]{aastex}
 
\newcommand{\etal}{\mbox{et~al.}}

\newcommand{\Msun}{\mbox{${\bf M_{\odot} }$}}

\def\deg      {{\ifmmode^\circ\else$^\circ$\fi}} 

 \shorttitle{The Stellar Content of the {\sl COSMOS} Field}
 \shortauthors{Robin et al.}
 
\begin{document}
 
 
 \title{The Stellar Content of the {\sl COSMOS} Field as derived from morphological and SED based Star/Galaxy Separation \altaffilmark{1} }
 

%
%
%
 \author{ Annie C. Robin \altaffilmark{43},
R. Michael Rich\altaffilmark{23},
H. Aussel\altaffilmark{4,20},
P. Capak\altaffilmark{1},
L. A. M. Tasca\altaffilmark{15},
K. Jahnke\altaffilmark{25},
Y. Kakazu\altaffilmark{4},
J-P. Kneib\altaffilmark{15},
A. Koekemoer\altaffilmark{5},
Alexie C. Leauthaud\altaffilmark{15},
S. Lilly\altaffilmark{8},
B. Mobasher\altaffilmark{5},
N. Scoville\altaffilmark{1,2},
Y. Taniguchi\altaffilmark{41},
D. J. Thompson\altaffilmark{33,42}}

 
\altaffiltext{$\star$}{Based on observations with the NASA/ESA {\em
Hubble Space Telescope}, obtained at the Space Telescope Science
Institute, which is operated by AURA Inc, under NASA contract NAS
5-26555; also based on data collected at : the Subaru Telescope, which is operated by
the National Astronomical Observatory of Japan; the XMM-Newton, an ESA science mission with
instruments and contributions directly funded by ESA Member States and NASA; the European Southern Observatory under Large Program 175.A-0839, Chile; Kitt Peak National Observatory, Cerro Tololo Inter-American
Observatory, and the National Optical Astronomy Observatory, which are
operated by the Association of Universities for Research in Astronomy, Inc.
(AURA) under cooperative agreement with the National Science Foundation; 
the National Radio Astronomy Observatory which is a facility of the National Science 
Foundation operated under cooperative agreement by Associated Universities, Inc ; 
and and the Canada-France-Hawaii Telescope with MegaPrime/MegaCam operated as a
joint project by the CFHT Corporation, CEA/DAPNIA, the National Research
Council of Canada, the Canadian Astronomy Data Centre, the Centre National
de la Recherche Scientifique de France, TERAPIX and the University of
Hawaii.}  
\altaffiltext{1}{California Institute of Technology, MC 105-24, 1200 East
California Boulevard, Pasadena, CA 91125}
\altaffiltext{2}{Visiting Astronomer, Univ. Hawaii, 2680 Woodlawn Dr., Honolulu, HI, 96822}
%
%
\altaffiltext{4}{Institute for Astronomy, 2680 Woodlawn Dr., University of Hawaii, Honolulu, Hawaii, 96822}
\altaffiltext{5}{Space Telescope Science Institute, 3700 San Martin
Drive, Baltimore, MD 21218}
%
%
%
\altaffiltext{8}{Department of Physics, ETH Zurich, CH-8093 Zurich, Switzerland}
%
%
%
%
%
%
%
\altaffiltext{15}{Laboratoire d'Astrophysique de Marseille, BP 8, Traverse
du Siphon, 13376 Marseille Cedex 12, France}
%
%
%
%
%
\altaffiltext{20}{Service d'Astrophysique, CEA/Saclay, 91191 Gif-sur-Yvette, France}
%
%
%
\altaffiltext{23}{Department of Physics and Astronomy, University of
California, Los Angeles, CA 90095}
%
%
\altaffiltext{25}{Max Planck Institut f\"ur Astronomie, K\"onigstuhl 17, Heidelberg, D-69117, Germany}
%
%
%
%
%
%
%
\altaffiltext{33}{Caltech Optical Observatories, MS 320-47, California Institute of Technology, Pasadena, CA 91125}
%
%
%
%
%
%
%
%
\altaffiltext{41}{Physics Department, Graduate School of Science, Ehime University, 2-5 Bunkyou, Matuyama, 790-8577, Japan}
\altaffiltext{42}{Large Binocular Telescope Observatory, University of Arizona, 933 N. Cherry Ave.
   Tucson, AZ  85721-0065,   USA}
\altaffiltext{43}{CNRS-UMR6091, Observatoire de Besan\c{c}on, Universit\'e de Franche-Comt\'e, F-25010 Besan\c{c}on cedex, France}

  
 \begin{abstract}

We report on the stellar content of the {\sl COSMOS} two degree field, as derived from 
a rigorous star/galaxy separation approach developed for using stellar sources to define the point spread function variation map used in a study of weak galaxy lensing. The catalog obtained in one filter from the ACS (Advanced Camera for Survey on the Hubble Space Telescope) is cross-identified with ground based multi-wavelength catalogs obtained using the {\sl SUPRIME CAM} instrument on the {\sl SUBARU} telescope, which makes possible detailed spectral energy distribution fitting in order to separate stars from QSOs and compact galaxies. The classification is reliable to magnitude $F_{814W}=24$ and the sample is complete even fainter.  We construct a color-magnitude diagram and color histograms and compare them with predictions of a standard model of population synthesis at $l=236.816\deg$, $b= 42.12 \deg$. We find features corresponding to the halo subdwarf main sequence turnoff, the thick disk, and the thin disk.   We propose improvements to the standard model that give a better fit : this data set provides constraints on the thick disk and spheroid density laws and on the IMF at low mass, although complementary lines of sight would help in lifting the degeneracy between model parameters as well as to mitigate any variations in the stellar populations.  The depth of this survey makes it possible explore the spheroid up to distances of $\sim 80$ kpc; we find no evidence of a sharp spheroid edge out to this distance, corresponding to a galactocentric radius of 83 kpc.
Beyond 60 kpc (the boundary of current surveys) we find a deficit
relative to the widely used $r^{-3}$ spheroid density law.
We identify a blue population of white dwarfs with counts that agree with model predictions.  We find a hint for a possible slight stellar overdensity at about 22-34 kpc but the data are not strong enough at present to claim detection of 
a stream feature in the halo.

\end{abstract}
 
 
\keywords{COSMOS field, stellar populations, deep fields, stellar halo, thick disk}
 


 \section{Introduction}

The {\sl COSMOS} survey field covers an unprecedented two degrees of contiguous imaging using the Advanced Camera for Surveys ({\sl ACS}) on board the
{\sl Hubble Space Telescope} ({\sl HST}). The complete survey is described in
\citet{sco07a,sco07b} in this volume while the photometric systems and catalogs are given
in \citet{Capak}, also in this volume.
The primary aim of the {\sl COSMOS} is to cover the largest contiguous field ever observed with HST imaging, enabling a study of galaxy evolution as a function of the full range of large scale structure context, over a redshift range of $0<z<1.5$ and ultimately to very high redshift.   The ultimate aim is for the {\sl COSMOS} survey to obtain comparably deep multiwavelength datasets covering the full field size;
deep data from X-ray to radio wavelengths have been obtained and 
results for many of these datasets are described in this volume. 

One of the central science goals of the ACS {\sl COSMOS} program is the construction of a map of weak gravitational lensing covering the full field \citep{rho06}.  The aim is to provide a map of the mass distribution in the field, which is to be compared with the distribution of luminous matter.  A byproduct of the weak lensing study is a rigorous selection of stellar sources across the full field.   While the weak lensing analysis identifies point sources, a significant number of these are extragalactic.   Lacking additional constraints like proper motions, we also use the multiwavelength  SED (Spectral Energy
Distribution) and a set of stellar and extragalactic template spectra to excise
as much of the extragalactic population as possible.
If additional {\sl HST} imaging can be obtained in the future, the proper motions of stars in this field will be of great interest and will enable a wide range of science, including an assessment of the white dwarf population in the Galactic spheroid.  The S-Cosmos dataset \citep{sanders07} 
also has the potential to add interesting red stars, such as L and T dwarfs.

The final field center  (RA=10h 00m 28.6s, +02d12m21s, J2000; 150.12\deg, +2.206\deg)
was driven by a wide range
of requirements, including low infrared background and avoidance in right ascension of
other extragalactic deep fields also requiring significant resources.   The field is coincident
with the D2 field of the CFHTLS, and is  $(l,b)=236.82\deg, +42.12\deg$, a relatively
low Galactic latitude for a deep field and therefore of interest to studies of the Galaxy.   While the {\sl COSMOS } field is coincident with D2,
we emphasize that our study here supersedes that of \citet{Schultheis2006}
which includes the D2 field, as our photometry is deeper and the
star/galaxy separation superior.  
Recent analysis of the SDSS imaging data \citep{bel-fld06} shows that {\sl COSMOS}
lies outside of the ``field of streams'' region and not coincident
with any other halo structures.

In this preliminary analysis, we use the Besancon galaxy model \citep{Robin2003} to compare the observed CMD with the candidate stellar populations that account for the observed populations. Our
study confirms that the combined morphology/SED technique recovers the stellar
content of the field very well and enables a probe of stellar population models to new depths.

The paper is organized as follows.  The following section discusses the observations and our
definition of the stellar sample.   Having obtained the sample, in Section 3 we consider the
observed stellar populations in the {\sl COSMOS} field, including the spheroid (3.1; 3.2)
the thick disk (3.3),  thin disk (3.4), white dwarfs (3.5), and a newly
discovered T dwarf candidate (3.6).  Section 4 gives our conclusions.

\section{Observations and sample selection}

The {\sl COSMOS} field consists of one orbit ACS images in the F814W (I band)
filter and ancilliary ground-based datasets described in Section 2.2 below.  The observations and analysis of the data are discussed in \citet{sco07a} and the overview of {\sl COSMOS} photometry is in \citet{Aussel07}.

In these images it is possible to detect point sources to $I_{AB}=27$ with $S/N\sim 10$, 
but secure star/galaxy separation is possible only at brighter magnitudes.
The selection of the stars from the sample has been made in two steps: first by morphology, second by SED fitting to remove remaining QSOs and compact galaxies contaminating the star sample.


\subsection{ Point source analysis by morphology }

Our decision to explore the stellar population was motivated by the success of the weak
lensing analysis of the {\sl COSMOS} HST images to select for stellar sources.  Precise characterization
of the point spread function (psf) is critical to undertaking the weak lensing analysis.  
We employ the  SEXTRACTOR  code \citep{Bertin} to accomplish this.  While the code produces
a continuous stellarity index parameter ranging from 0 (extended sources) to 1 (point sources),
\cite{Leauthaud} find the parameter to have two drawbacks.  First, the definition of the dividing line between stars and extended sources is ambiguous.  Second, the neural-network classifier used was 
trained with images obtained in ground-based seeing and consequently is (in principle) valid for a sample of profiles similar to the original training set, and less well
calibrated for HST images.  \cite{Leauthaud} explored the use of
this index nonetheless, but found the results hard to interpret. They therefore chose to use the 
SExtractor parameter {\tt MU\_MAX} (peak surface brightness above the background level) and
another parameter {\tt Rhl} which is based on the half-light radius.   Both approaches rely on point
sources following the psf while extended sources (e.g. galaxies) do not.   
Figure~1, reproduced from \cite{Leauthaud}, shows the results of this exercise.
Point sources
occupy a well-defined region in the {\tt MU\_MAX} vs {\tt MAG\_AUTO} plane.  {\tt MAG\_AUTO} is another
SEXTRACTOR output parameter that can be considered to be a dynamic aperture magnitude with the
aperture scaled to the half-light radius.  Experiments comparing the objects selected using
the {\tt Rhl} and {\tt MU\_MAX} parameters indicate excellent agreement.  However,
{\tt MU\_MAX} gives a tighter correlation for the stellar locus and a clear break at the magnitude
where stellar sources saturate {\tt MAG\_AUTO}=19.    The {\tt MU\_MAX} method also is used
to define artifacts, point sources more sharply peaked than the psf.  A visual inspection of these
sources confirm that this population is dominated nearly completely by hot pixels, residual
cosmic rays, and other artifacts; these are flagged and omitted from the stellar catalog.

Over the regime we are considering, the star/galaxy separation method has been tested for
completeness and is complete and reliable at the 90\% level to $I_{AB}=27.1$.   The analysis
of completeness is described in detail in \cite{Leauthaud}: in brief, Gaussian sources are
inserted into the ACS frames and then recovered.  In principle, the use of a real stellar psf would
be the preferred approach, but we do not push our analysis into the regime where completeness
is a concern.  An empirical approach to completeness is to compare star counts with numbers of
all sources reaching the limiting magnitude of F814W=30.  At F814W=27, the counts of
all sources plunge, and star counts also show a sharp decline.  By limiting our analysis to F814W=25,
we are working in a regime where completeness should be secure.

 \subsection{Multi-wavelength data}

The starting point is an ACS selected stellar objects catalog; this was constructed
from the full ACS catalog by 
extending the point source cylinder all the way to {\tt MU\_MAX}$=25$. 
The stellar catalog has then been correlated with the ground-based photometry 
catalog (the COSMOS photometric catalog dated 3 January 2006 \citet{Mobasher}) in
order to identify the colors of objects when possible. 
The catalog contains a total of 237998 objects of which 149679 are matched to 
the ground based catalog (and therefore have several colors).

The multi-color catalog \citep{Capak} consists of photometry 
measurements over a 3 arcsecond diameter apertures for deep 
$B_j, V_j, g+, r+, i+, z+$ Subaru data, $u*, i'$ Canada-French Hawaii
Telescope bands (CFHT), $ugriz$ photometry from the Sloan 
Digital Sky Survey (SDSS), $K_s$ magnitude from KPNO/CTIO, and $F814W$ 
HST/ACS magnitude.
Here after we refer to band $I_{ACS}$ for the photometry in F814W filter of the ACS camera, to $r_{subaru}$ for $r$ magnitude from SuprimeCam at Subaru Telescope, to
$u, g, r, i, z $ for the photometric system of CFHT (megacam camera), and to $K_{kpno}$ for K band magnitude from KPNO/CTIO, all magnitudes being in the AB system.

\subsection{Photometric redshifts}

In order to identify probable stars among stellar candidates selected by morphology, and to make the best use of the multi-wavelength data, we have used the spectral energy distribution (SED) fitting
method derived from the photometric redshift fitting methods, as follows:

We use the code {\tt Le Phare}\footnotemark{1} which 
is based on the $\chi^2$ fitting procedure by comparing the
observed magnitudes with the magnitudes predicted from a library of
spectral energy distributions.\footnotetext{
The code {\tt Le Phare} by S. Arnouts and O. Ilbert is available at the URL: http://www.oamp.fr/arnouts/LE\_PHARE.html.} 
The method and details
are described in \citet{arnouts99}. A calibration
method can be applied by using the spectroscopic redshift sample as
training set; this method is described in \citet{ilbert06}.

Our primary galaxy template set are the four Coleman, Wu and Weedman (CWW)
observed spectra: Elliptical (Ell), spirals (Sbc, Scd) and irregular (Irr) \citep{cww80}.
These templates are linearly extrapolated into ultraviolet ($\lambda < 2000 \AA$)
and near-infra-red wavelengths using the GISSEL synthetic models \citep{bc03}. 
For spectral types later than Sbc a reddening has been introduced.
The template optimisation method is described in \citet{ilbert06}.

The stellar template set is a combination of BaSeL2.2 semi-empirical spectra \citep{Lejeune97} for high temperature stars and the NextGen atmosphere models from \cite{Allard97} for cool main sequence stars.  We prefer use of the templates because 
they span a greater range
in metallicity and color than the standard star library \citep{pic98} 
and allow us to be more complete in our star catalog.
Note that this point is important for understanding 
the selection of the objects.

For the QSOs we use synthetic spectra associated with the {\it Le Phare} code, with the addition of a QSO template from  \cite{Cristiani}.

For each object we compute simultaneously the $\chi^2$ for the galaxy library,
the $\chi^2_q$ for the QSO library and the $\chi^2_s$ for the star library.
If the condition $\chi^2 - \chi^2_s >0$ holds the object is flagged as a star. 
The method applied allows us to reach an accuracy for  the redshift $z$ 
$\sigma_{\Delta z}/ (1+z_s) = 0.031$ with $\eta = 1.0 \%$ the fraction of catastrophic
errors, defined as $\Delta z /(1+z_s) >0.15$ \citep{Mobasher}.

Among the 16055 stellar candidates at $i_{ACS} <25$ for which multi-wavelength data are available, we have identified 11403 objects classified as stars,  2296 objects
classified as QSOs and 2356 objects in the galaxy category.

In order to test the completeness of the star counts obtained with the SED fitting we have attempted to use a relaxed $\chi^2$ test where more objects are classified as stars. This is at the cost of having a greater contamination by extragalactic compact objects. Then the $i_{ACS}-K_{kpno}$ color can be used to separate the stars from galaxies. Figure~2 
 shows, in the upper panel, the $r_{subaru}-i_{ACS}$ versus $i_{ACS}-K_{kpno}$ diagram on which a stellar sequence is clearly visible and contamination by non stellar objects is clear for the reddest $i_{ACS}-K_{kpno}$ values.  We have plotted a line for separating these objects which is defined by $i_{ACS}-K_{kpno}={seq}+0.5$ and $seq$ is a rough fit to the stellar sequence. Then we select those objects below the sequence and the resulting color-magnitude diagram is shown in the middle panel of the figure. At the bottom we show the objects which have been classified as galaxies from this process. The rejected objects lie in a region where galaxies are expected. The resulting star counts are shown in figure~3.  
The solid line shows the objects classified as stars with the relaxed selection from the SED. The dashed line is the objects classified
as stars from the strict SED selection. The dotted line gives the star counts with the relaxed SED selection and the $i_{ACS}-K_{kpno}$ selection explained above. The analysis shows that up to magnitude 23.75, both methods of strict selection are in very good agreement. Fainter than this, the "relaxed+($i_{ACS}-K_{kpno}$)" selection leaves many galaxies due to large errors on $k$ at faint magnitude, the data being less sensitive at $K_{kpno}$. Finally, we choose to keep the SED based selection, which is reliable to mag $i_{ACS}=24$, but this
becomes progressively less reliable at faint magnitudes.

\section{Stellar populations in the COSMOS field}



In the CMD obtained from the strict SED selection shown in the middle panel of figure~2, one can identify 4 main features:
 
 \begin{itemize}
 \item a blue sequence at $r_{subaru}-I_{ACS}$~=~0.2 in the range 18$<I_{ACS}<$23.5. We interpret this feature as the spheroid
population at the main sequence turnoff.

\item a prominent red sequence corresponding to red dwarfs from the disk population. The sequence is well defined up to magnitude 23. However at fainter magnitudes the number of such objects decreases
significantly, either due to the lower sensitivity of Subaru $r$ compared with ACS $I$, or due to a decrease in the number of low mass dwarfs in the disk population. This is analysed in more detail in section 3.4.

\item starting at about $I_{ACS}=21$ a middle sequence which can be attributed to thick disk and/or spheroid subdwarfs. This point will be investigated using the Galaxy model simulations in sections 3.2 and 3.3.
 
 \item The classification is unreliable at $I_{ACS}>26$, as seen above. 

\end{itemize}

In order to pursue an initial quantitative analysis of the stellar populations,
it is useful to simulate the expected stellar populations in this field. 
For this purpose we have used the Besancon Galaxy Model (here after BGM).  A complete description of the model can be found in \citet{Robin2003}. A succint description follows.

Hess diagrams  (a color-magnitude diagram with number density) are computed from evolutionary tracks and constrained by Hipparcos data. The main ingredients are: constant star formation rate for the disk from 0 to 10 Gyr, starbursts at 10, 11 and 14 Gyr for the bulge, thick disk and spheroid populations respectively), a particular choice of IMF, vertical and radial stellar density laws constrained either by dynamical consistency (disk) or by remote star count fitting (thick disk and spheroid). Apparent magnitude and colors in various photometric systems are computed using the BaSeL synthetic photometry database \citep{Lejeune97}. 
Generally magnitudes and colors are computed for the standard Johnson-Cousins photometric system, and scaled to the Vega system. 
%
%
%
 
However, the colors which have been observed are closer to the SDSS $ugriz$ system than to the Johnson-Cousins system. In particular, the $g$ band covers both the B and V Johnson bands, making a color equation from $g$ to B or V uncertain. Hence a comparison is better made in a system close to the observed one. The BGM has been calibrated to produce simulations in the Megacam camera photometric system \citep{Schultheis2006}. We use this version of the model here.

The data must also be converted to the same photometric system. Using the star templates already employed for SED fitting, color transformations have then been computed from the ACS F814W and Subaru bands into the Megacam bands $u^*g^\prime r^\prime i^\prime z^\prime $ in order to be able to compare with model simulations. 
With these theoretical equations, colors of the observed sample in the Megacam photometric system have been computed and compared with simulations. 
In the following we refere to this system by using magnitudes $i$ and color $r-i$, both systems being close for these bands.

Figure~4 
shows this observed CMD in the Megacam system $r-i$ versus $i$ and figure~5 
the simulation with the standard model. The observed sample contains fewer stars than shown in figure~2, in the CMD with $I_{ACS}$ and $r_{Subaru}$, due to the omission of stars with uncertain conversion to the Megacam system (bad fit) especially at faint magnitudes. Colors in the model are for different populations (red: thin disk, green: thick disk, blue: halo). The populations are
rather well separated in the CMD. 

While color-magnitude diagrams are useful to determine the location of various populations, color histograms help in quantifying how well the density predicted in the model agrees with the data.
Figure~6 
shows comparisons of histograms of $r-i$  colors
in various magnitude ranges between the data and the model (data in solid line, two models in dashed and dotted lines). The range in $i$ magnitude is given in the upper right corner of each panel.
The populations are rather well identified by their $r-i$ color as can be shown using the model: 
figure~7 
shows the same histogram for the model
only but with a color code for each population superimposed: red: thin disk, green: thick disk, blue:spheroid.

From the CMD and color histograms stellar population features are clearly identified. In the following we study each feature in more detail and propose modifications to some model parameters to better fit these features.

\subsection{The spheroid turnoff and density law}
\label{spheroid}

Figure~5 
shows the simulated CMD from the standard BGM.
Compared to figure~4, 
we find that the principle sequences 
of the CMD are reproduced reasonably well; nevertheless, we suspect that the model requires
additional parameters to reproduce fully the CMD.

The first interesting feature concerns
the blue boundary of the spheroid main sequence turnoff that
is well defined at 
$0.0<r-i<0.3$, in both model and data. 

For a population having a single age distribution and homogeneous metallicity, the blue boundary of this
locus should have the same color at all magnitudes. But we do see distinct variation in this blue boundary as we sample the spheroid turnoff point at a range of distances.  Note that at $i<19$ 
the images are saturated.   

It should be noted that the turnoff sequence of the spheroid occupies a well
defined locus in the simulation, with a trajectory that reddens monotonically
with increasing apparent magnitude.  This is not the case for the data, where
the blue boundary is seen to follow a curving trajectory
with apparent magnitude, a behavior that is not reproduced in our standard model.  

The postulated turnoff population shows little variation in density to $i=22-23$; mag 23 corresponds to about 80 kpc
and is not detected fainter than this, as this may well
correspond to the boundary of the spheroid.
Some of the slight shift to redder colors of this sequence almost certainly reflects the contribution of
fainter (redder) spheroid main sequence stars along the line of sight.
At fainter magnitude it becomes less well marked, due to the lack of stars at these distances. 

As these are spheroid turnoff stars, they may be used to constrain the
spheroid density law, subject to uncertainties in distance.
From \cite{Bergbush} the absolute magnitude at
the turnoff varies slightly with the assumed age of the spheroid, from $M_V$=+3.9 for a 13 Gyr isochrone to +4.0 for 14 Gyr. However, a larger uncertainty arises because a range of only 0.01 magnitude in color along the isochrone covers a range about $\pm 0.25$ in absolute magnitude near the turnoff. Hence the uncertainty in distance for turnoff stars is rough 12\% for a 1\% uncertainty in color.

Assuming that turnoff stars are at $M_V$=+4.0, or $M_i=3.72$, the extinction being about E(B-V)=0.06 in the field  \citep{Schlegel} the distance of those stars are in parsecs:

\[ \mbox{distance} = 10^{0.2 \times (i-M_i+5.) } = 10 ^{0.2 \times (i+1.28)}\]

This quantity has been computed from the model and data catalogue for stars selected by 
0.$<r-i<0.3$ corresponding to the blue peak. 

In order to compare the density law implied by the observations with
that of the model, Figure~8 
shows the comparison of histograms of this
distance estimator in both catalogs selected as 0.$<r-i<$0.3. 
The shape of the spheroid density law is assumed to follow a power law as:
\[ \rho \propto \alpha^{-n} \]

with $\alpha = { R^2 + z^2/\epsilon^2}^{1/2} $  
where $n$ is the power law exponent, $R$ and $z$ are galactocentric cylindrical coordinates and $\epsilon$ is the axis ratio taken equal to 0.76 \citep{Robin2000}.
 
Two models are presented with two different power law exponents for the halo density law: 2.44 as in \cite{Robin2000} and 3. 
The latter value better fits the data out to a distance of 70 kpc. More distant still, either the exponent is even larger or we start
to see the edge of the spheroid; photometric errors
are significant in the fainter parts of the CMD.   The density law actually depends on
several parameters, such as the exponent and the flattening, as shown in \cite{Robin2000}.  Hence a fit of the density law along only one sight line is not sufficient to disentangle the correlation between both parameters.
Ideally, it would be preferable to sample a large number of different sightlines at
comparable depth and area to the {\sl COSMOS} field. 

The distance distribution looks quite smooth and monotonic, apart from a slight enhancement around distance 25-35 kpc that might be related to the curving of the turnoff locus seen in the CMD. We detect this feature at only a $2\sigma$ (Poisson counts) level. 
In order to check whether the feature at i$\approx$  21-22 and  $r-i \approx$ 0.1 corresponds to a real
overdensity of objects in the halo at distances 22-34 kpc, we have simulated such a population
with equivalent characteristics as the spheroid (age and metallicity) and extending between 22 and 34 kpc. Figure~9 
shows the CMD of the data (in the region of the spheroid) compared with simulations without and with such a extra population. The model including the
extra population seems to best fit the data at the turnoff position. However farther down the main sequence the number of stars in this simulated extra population appears too large at roughly  i$>$23 and $0.2<r-i<0.5$. In the data, the main sequence of this possible extra population appears to vanish. Alternatives would be to change the age, the IMF or the metallicity of the population in the simulation but it is difficult to imagine any prescription that would truncate the faint main sequence, as that implies an unphysical truncation of the mass function. However, further exploration must be deferred until we have a deeper dataset, either through second epoch HST imaging to 
give proper motions or by using the BzK method, when significantly deeper K band imaging becomes available.

\subsubsection{The Distant Spheroid}

As mentioned earlier, Figure 8 shows that our data is not consistent with our
having observed a boundary to the halo.
The most secure present constraints
on the extent of the halo are from counts of RR Lyrae stars \citep{ivezic2004, Vivas}; however, RR Lyraes arise only in
stellar populations older than 10 Gyr while the turnoff population traces the complete population regardless of age.
On the other hand, the distances of RR Lyrae stars can be determined with far greater precision than is possible
for main sequence turnoff stars.

The distribution of the turnoff stars does not seem to show any cutoff or break in the power law out to 80 kpc, where
the data are limited by the depth of our analysis. Deeper data in this field exist but they are contaminated
by galaxies. Spectroscopic data or deeper K band photometry (for a better selection of extragalactic objects from SEDs, the efficiency of the SED fitting being limited by the depth of the near-infrared data)
would be needed to constrain the boundary of the spheroid. Again, proper motion data would be
helpful here to separate halo stars from galaxies, however the required accuracy (of the order of 0.5 mas/yr for a halo star at 80 kpc) would be difficult to obtain.

Is our finding of an extended halo corroborated by other observations?
The most extensive counts of halo RR Lyrae stars are derived from the  SDSS \citep{ivezic2004}.  
They find a constant slope $r^{-3}$ profile density law to distances of at least 60 kpc with
no sign of a boundary; these
findings are confirmed to 40 kpc in the Quest survey \citep{Vivas}, which spans 165\deg in right
ascension.   However, one system, the metal poor globular cluster NGC 2419, is at
a distance of 84 kpc \citep{harris97} and roughly in the direction of the {\sl COSMOS}
field.  Debris (likely arising from the Sgr dwarf) have been observed at 80 kpc from
the Galactic center at $(l,b)=350,50$ \citep{dohm01} while theoretical models
of the Sgr dwarf debris stream that match the SDSS data  \citep{fel06} predict
that debris should be found out to 60 kpc from the Galactic Center.  Considering also the extensive
discoveries of new streams from studies of SDSS \citep{bel-fld06} one may conclude
that our findings are in fact consistent with a wide range of other evidence concerning the distant halo.

In fact, the outermost boundary of the halo might well be irregular, reflecting the presence of streams and
infalling satellites.  As the SDSS data have been analyzed to a greater extent, the halo of the Galaxy is
beginning more and more to resemble that of M31 \citep{ferg02}, consisting of infall streams
and large, irregular stellar concentrations
(Belokurov et al. 2006; Belokurov 2006, private communication). 
In standard halo formation models in a $\Lambda$CDM universe, these outer regions
should retain a record of accretion events longer than the inner parts of the halo, the dynamical relaxation time reaching several Gyr (according to simulations from \cite{Bullock2005}).  If the outer boundary of the halo is defined by these multiple
 accretion events, it is conceivable that the derived density law and boundary 
 may vary significantly among different lines of sight.
 As with many of our studies here, it will be important to explore this issue with
similarly deep datasets in other directions. An approach like the one used by \cite{Gould} would be  appropriate, with a large coverage in longitude and latitude, but much larger statistics in every direction, giving access to the detailed structure of the spheroid.

\subsection{Spheroid main sequence stars}

A second feature is the density in the CMD of faint stars at $i>24$ occupying the faint extension of the
putative spheroid turnoff sequence: the feature appears more dense in the data than in the
simulation.  In an early study of this dataset, the distinct appearance of this sequence caused us
to consider the possibility that we had discovered the main sequence population of a stream or
dwarf galaxy residing in the halo. However it would not produce the expected feature shape, as seen from simulations in figure~9. 
While the most attractive explanation for this population is that we observe the halo subdwarfs at a range
of distances, the simulation still falls short in predicting the numbers of these stars at 0.6$<r-i<1.4$.   
One possible route to an improved fit is to 
increase the IMF slope at low masses for all the population components in the
model. 
This has been done in the alternate model which appears
in figure~10 
where the IMF slope is changed from $\alpha=0.5$ to $\alpha=1.5$. The simulations are clearly sensitive to this parameter.
However the value $\alpha=1.5$ might still be an underestimate, if that part of the CMD is less complete (due to the limiting $r_{subaru}$ band of the ground-based observations). As explained in \cite{Schultheis2006} and \cite{Larson} the IMF of old populations
with a formation redshift larger than 2 may well reflect varying environmental
conditions (e.g. metallicity, radiation field, etc.) that affect the Jeans mass.

\subsection{Thick disk population}

A third feature in the CMD is found in the central region at colors $0.5<r-i<1.5$. The thick disk as
simulated in the model appears too prominent and is seen at magnitudes too bright
compared to the data. The thick disk is modeled by a truncated exponential.
This truncated exponential is a modification at short distances of the exponential in order to have a differentiable function in the plane at $z$=0. The function is exponential at distances larger than 400~pc,
 with 
a scale height of 800 pc and a local density normalised to 4.6\% of the thin disk \citep{Reyle2001}. 
These parameters have been derived from a large dataset that is not as deep as the
present sample but covering a wide range in Galactic longitude and latitude.  We have noted in the latter paper a strong degeneracy between the scale height and the local density. 
The choice of the scale height determines the position of the thick disk sequence in both
apparent magnitude and color.
In figure~10 
we propose an alternate model
where the thick disk scale height is 1200 pc and the local density of 1.15\% of the thin disk. With these
parameters the density of intermediate color stars is lower and better in agreement with the data.
A scale height of 1000 pc is marginaly compatible with the data at the 3 sigma level and the local density
is constrained at about 30\% daccuracy. The absolute magnitude range which dominates the solution is $9<M_V<11$, that is near the peak of the luminosity function.
However we cannot make conclusions with respect to the scale height value of the thick disk 
from this sample alone because the solution also would depends on the assumed scale length; analysis of other sightlines is required. A larger study
including deep data towards several directions would be necessary in order to constrain the thick disk model.  Moreover it may occur that the parametrization (double exponential, rather than sech squared, or other shape) may not be appropriate for all directions. We only see here that the thick disk density law assumed in the standard model is not optimal for this field.

\subsection{Thin disk population}

Finally the reddest part of the CMD contains red dwarfs from the thin disk population. A detailed comparison of this sequence between the observations and the simulation of the standard model (figure~5)
shows that the position of the sequence in $r-i$ is not well
placed in the simulation at $i>$22 and that the number of very red stars $r-i>2$ seems to be overestimated. The color of this sequence is modeled using NextGen stellar atmospheres \cite{Allard97} convolved with our filter bandpasses. The same spectra and filter definition have been used for recomputing the observed colors in the Megacam system and in the Galaxy model. Hence the discrepancy could come from uncertainties in the modelling of the atmosphere of very cool dwarfs. 
Figure~11 
shows a color-color diagram of the present sample compared with NextGen model colors for dwarfs of temperature below 4400 K at solar metallicity and sub-solar metallicity. The $r-i$ color predicted by NextGen model appears too red for a given $i-z$, from $r-i>1.5$ corresponding to an effective temperature of 3000 K. For solar metallicity the effect is not so strong but it increases significantly at lower
metallicity.  The mean metallicity of the thin disc in about -0.2, giving a shift in $r-i$ of about 0.3 magnitude at $r-i=2$ between the model and the data. This is also noticeable in $g-r$ (too red in model atmosphere compared with this data set). 


We notice also an overestimate of the number of red stars predicted at $i>23$ in the standard model.
This could be related to the IMF slope used here, or due to the inadequacy of the model atmospheres.
 \cite{Schultheis2006} have used CFHTLS data (a deep survey of four
 one square degree fields in different directions) to constrain the IMF of
 the disk population at low mass.  Their data set was limited to magnitude 21 due to the difficulty of star/galaxy separation in the ground based images. They varied the IMF slope (multiple slope power law IMF) for these data at $r-i>$1.4 and
 conclude that a rather steep slope of $\alpha=2.5$ (notation: $\Phi(m) \propto m^{-\alpha}$) was necessary to fit this data set in 3 fields at longitudes 96, 172 and 236 $\deg$ and latitudes between 42 and 59 $\deg$. These data are much more sensitive to the range of mass from 0.3 to 0.1 solar mass and consider distances above the plane smaller than 400 pc.   
In contrast, our {\sl COSMOS} sample is much deeper, with star-galaxy separation in the red reaching $I\approx 24$, corresponding to the hydrogen burning limit. The number of very low mass stars is some 10 times greater than in the CFHTLS data due to this faint limiting magnitude, permitting us to study a much greater volume. 
The average distance of the sample is about 700 pc above the plane. This sample should constrain somewhat better the very low mass IMF and model simulations show that the extrapolation of the IMF with $\alpha=2.5$ used here following \citet{Schultheis2006} can clearly be ruled out, assuming that the model atmospheric colors are reliable.
In order to test this assumption, we have used an IMF slope of $\alpha=1.0$ at $m<0.2$\Msun\ in the alternate model presented in figure~10. 
This model is a better fit to the observed ACS CMD, but suffers from the
previously mentioned issue of being a single sightline in a halo that may well
have its own ``cosmic variance'' issues of subclumps etc.
Moreover, the result depends strongly on the model atmospheres adopted. The transformation from  $r-i$ color to luminosity and then mass relies on those models. Since we have shown above the models are uncertain at temperature below 3000 K, the latter conclusion should be revisited when better atmosphere models will be available. 

\subsection{White dwarfs}

Blueward of the spheroid turnoff locus lie candidate white dwarfs (WD).  The
area is potentially contaminated by stars with large
photometric errors. In figure~12
the CMD from data and models are shown. The white dwarf selection is made with -0.5$<$r-i$<$-0.08 to minimize the contamination by outliers of the spheroid turnoff.
The lower panel shows a comparison between white dwarf counts in the data assuming this color
selection (solid line), together with model predictions from the three populations and the total (red dotted line). The thin disk makes the major part of the counts. At magnitude fainter than 22.5 the number of
blue objects in the data starts to grow faster than expected, as clearly visible in the histogram of figure~12. This could be either due to a contamination from the QSOs, expected to grow with magnitude, or due to a larger than expected contribution of halo white dwarfs. The latter conclusion seems improbable. However if a few ancient white dwarfs would be present in the field, their high proper motions would easily distinguish them from quasars. 

A few objects bluer than $r-i=-0.5$ are present in the field, a color that is difficult for
stellar objects to reach.  We suspect that these are either extragalactic or
spurious.  The numbers of these objects are so small that one can imagine targeting
these in the course of large spectroscopic surveys. 

\subsection{A Brown Dwarf, and Future Prospects}

The possibility for the COSMOS survey to identify interesting and unusual stars is in its infancy.  Deeper $K$ band
imagery is in progress, and second epoch proper motion data are essential.  Already, however, searches for
high redshift objects have turned up one interesting result (Figure~13).  This T dwarf candidate (based on the
9400\AA ~water absorption) was identified as a promising high redshift candidate ($I_{ACS}-z_{subaru}>2, z_{subaru}-K_{kpno}<1$) and no
$u$ through $r$ counterpart.  We expect large numbers of other interesting objects, including halo white dwarfs,
to be identified in this field in the future, with substantially greater numbers if proper motion data 
become available.

\section{Conclusions}

We have used a primary morphological selection cut applied to the
HST/ACS data, followed by an SED-based cut, to select the stellar
population of  the {\sl COSMOS} field.  
We use the Besancon Galaxy model to generate model color-magnitude diagrams for the
stellar populations observed in the COSMOS field at $ l=238.8, b=42.1$.
We draw the following conclusions from this sample :

\begin{itemize}
\item The spheroid population is well detected to a distance of roughly 80-100 kpc. 
The spheroid is consistent with a smooth distribution, apart from a slight
$(2\sigma)$ enhancement in density at distances
between 22 and 34 kpc.   The possible overdensity may correspond to a halo tidal
stream or similar feature; further work is needed to explore the nature of this feature.  
We are unable to explain this excess with a plausible stellar population: we can fit the main 
sequence turnoff region but the population would need to have an unphysical decline in the main
sequence mass function, to fit the counts at fainter magnitudes.
 
\item The data are in favor of a spheroid having a density law modeled as a power law with an exponent of 3 or greater, with an axis ratio of 0.76. However it is not possible from only a
single line of sight to confidently constrain different parameters of the halo density like the power law exponent and the axis ratio.  We note that our observation of a relatively extended halo
population may pertain only to this particular field, with the halo perhaps having 
a somewhat irregular outer structure. 

\item The luminosity function of the spheroid main sequence could certainly be constrained with this data set, however it is difficult to separate from the thick disk population. A few other directions at different latitudes would help to break the degeneracy; further, it would help to have proper motion data.

\item The thick disk population is present in the middle of the color-magnitude diagram. The standard model simulation shows this structure but does not fit satisfactorily the position of the
thick disk in the observed diagram. A better fit is obtained with a thick disk scale height of 1200 pc$\pm 200$ and a local normalisation of 1.15\% of the thin disc, although this result is at odds with previous results obtained from wide data sets in many lines of sight. We shall investigate on a larger scale whether the thick disk shows  a smooth structure as simulated here, rather than clumpiness or at least variations from one line of sight to the other, as expected in a scenario of formation of thick disk as relics of accretion of nearby dwarf galaxies.

\item The thin disk low mass stars appear to be poorly simulated using the NextGen atmosphere models at effective temperature below 3000K. The COSMOS data would give good constraints on the IMF at very low masses because it reaches a large volume of space up to 700 pc above the plane for stars  down to the hydrogen burning limit. We shall investigate this point in the near future with alternate model atmospheres.  Deep $K$ band observations, the S-Cosmos deep Spitzer imagery, and proper motions, 
may shed light on what will be a significant survey of the lower main sequence and brown dwarfs in this field.

\item The white dwarf sequence is well distinguished blueward to the spheroid turnoff up to $I\approx 25$. The luminosity function is well in agreement with the simulation, hence the quasar contamination should be negligible up to this magnitude.  Our model indicates that this sample is dominated by thin disk white dwarfs but the model simulation shows that a handful of thick disk and halo white dwarfs might be present. We shall look forward for proper motions in this field, which would allow to identify them and sort them out from a possible residual quasar  contamination.
\end{itemize}

The {\sl COSMOS} field is presently the focus of an international campaign
to obtain a uniformly deep multiwavelength dataset.  While the emphasis is
on the galaxy evolution science, the dataset will provide one of, if not the,
best deep single multicolor dataset obtained for faint stars.  It will be of great value to
exploit this dataset to learn about the structure and stellar content of the
Milky Way as well as the properties of the faintest stars.

 \acknowledgments
 
 The HST COSMOS Treasury program was supported through NASA grant
HST-GO-09822. We wish to thank Tony Roman, Denise Taylor, and David 
 Soderblom for their assistance in planning and scheduling of the extensive COSMOS 
 observations.
 We gratefully acknowledge the contributions of the entire COSMOS colaboration
 consisting of more than 70 scientists. 
 More information on the COSMOS survey is available at \\
  {\bf \url{http://www.astro.caltech.edu/\~ cosmos}}. It is a pleasure the 
 acknowledge the excellent services provided by the NASA IPAC/IRSA 
 staff (Anastasia Laity, Anastasia Alexov, Bruce Berriman and John Good) 
 in providing online archive and server capabilities for the COSMOS datasets.
 The COSMOS Science meeting in May 2005 was supported in part by 
 the NSF through grant OISE-0456439. KJ acknowledges support by the German DFG under grant SCHI 536/3-1.

 
 
 {\it Facility:} \facility{HST (ACS, NICMOS, WFPC2)}

 \clearpage
 
 
 
\begin{figure}
\label{star-gal-separation}
\plotone{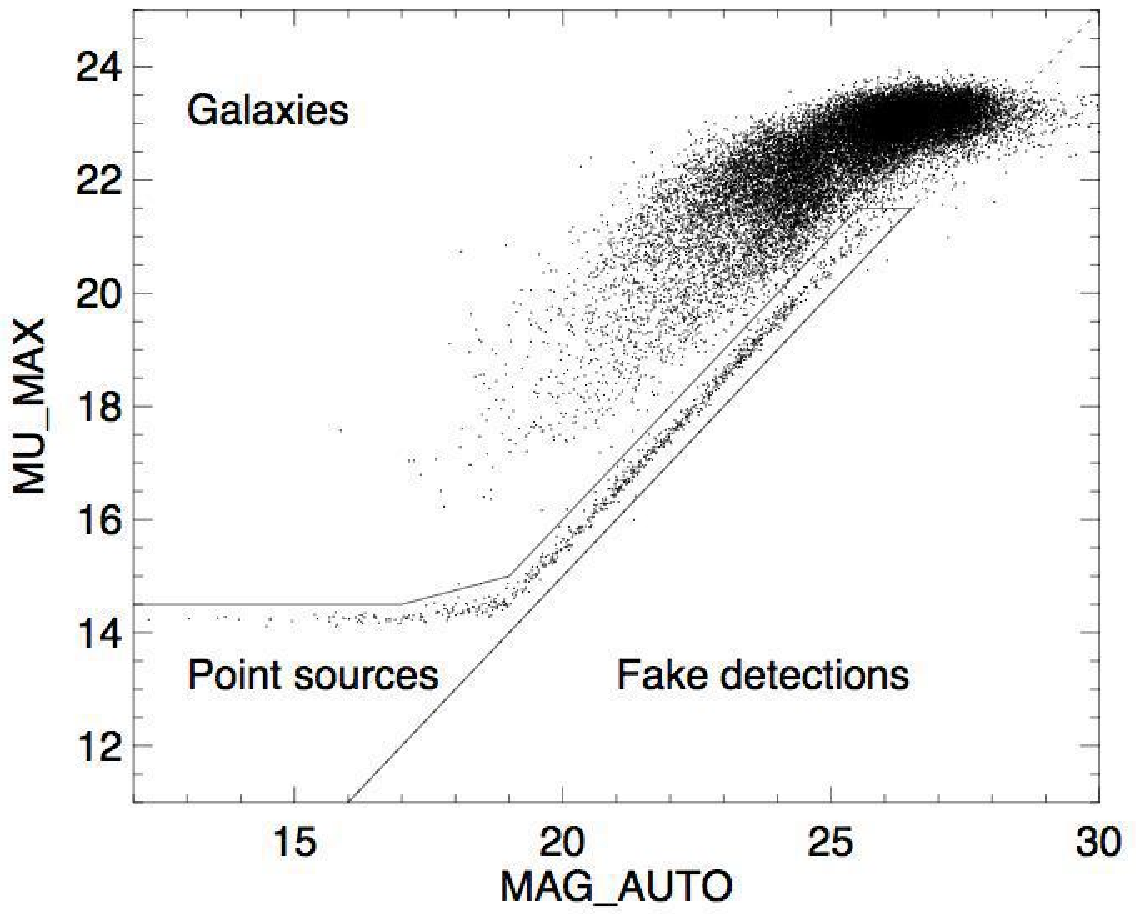}
\caption{ Classification of point sources, galaxies, and artifacts in the {\tt MU\_MAX} / {\tt MAG\_AUTO} plane; these classifications result from the application of the {\tt SEXTRACTOR} code 
to the images (see text and \citet{Leauthaud}).   Point sources follow the psf and lie in the polygonal region bounded by the solid lines at lower left.  Objects more sharply peaked than the 
psf lie to the right of the dashed line and are considered to be spurious artifacts.  For clarity,
we illustrate only a 2\% random selection of all objects in this plot.}
\end{figure}

\clearpage



\begin{figure}
\label{iksel}
\plotone{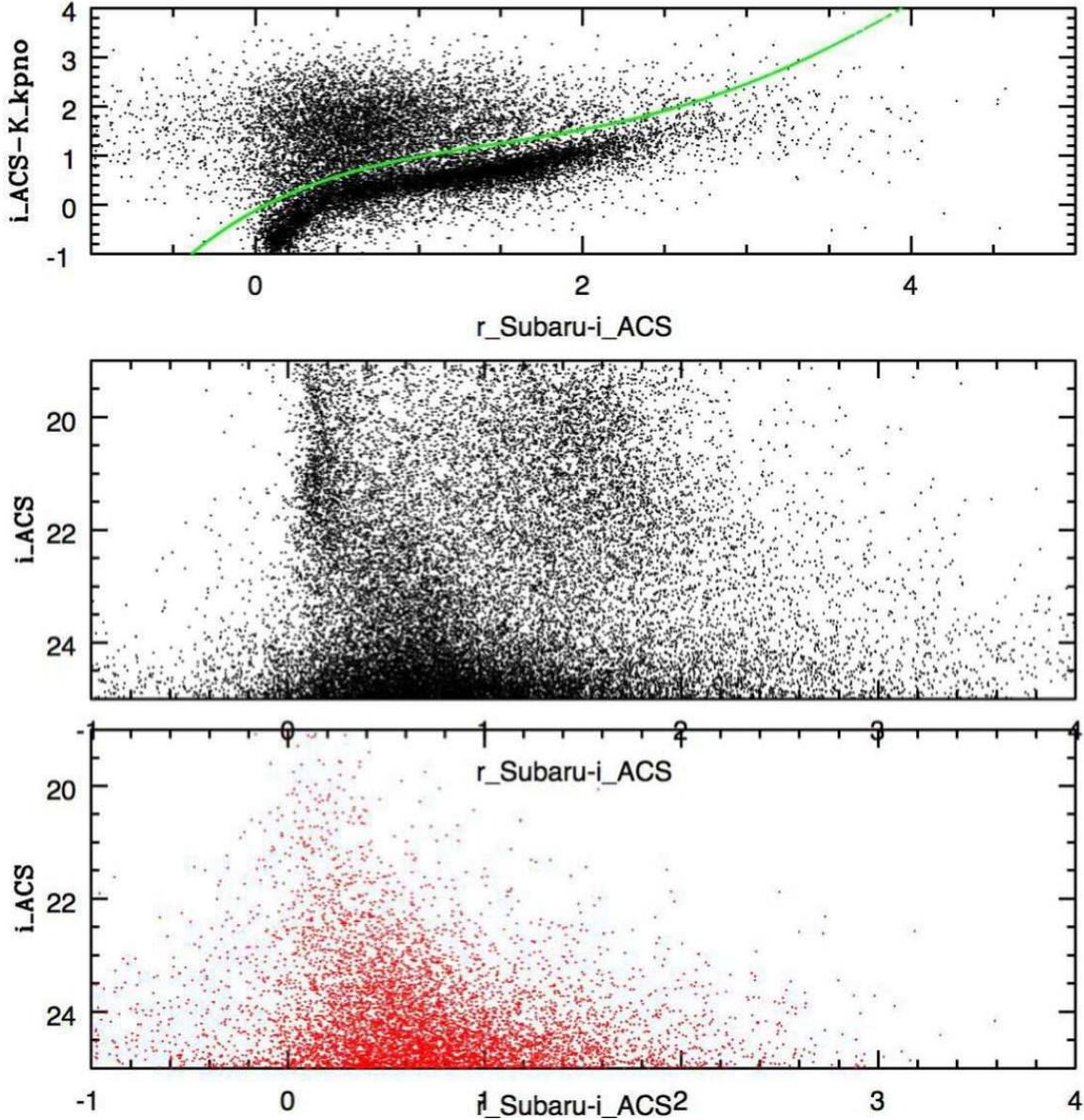}
\caption{Selection of stellar objects by a relaxed SED fitting criterion. Top panel: $r-i$ versus $i-k$ diagram
of objects classified as stars. A well defined stellar sequence is visible together with a sequence of galaxies at larger $ i-k$ for a given $r-i$. In the middle panel objects which are below the line in the upper panel are shown in a CMD. At the bottom the objects selected as galaxies (above the line in the upper panel) are shown in a CMD.}
\end{figure}
\clearpage

\begin{figure}
\label{clean-counts}
\plotone{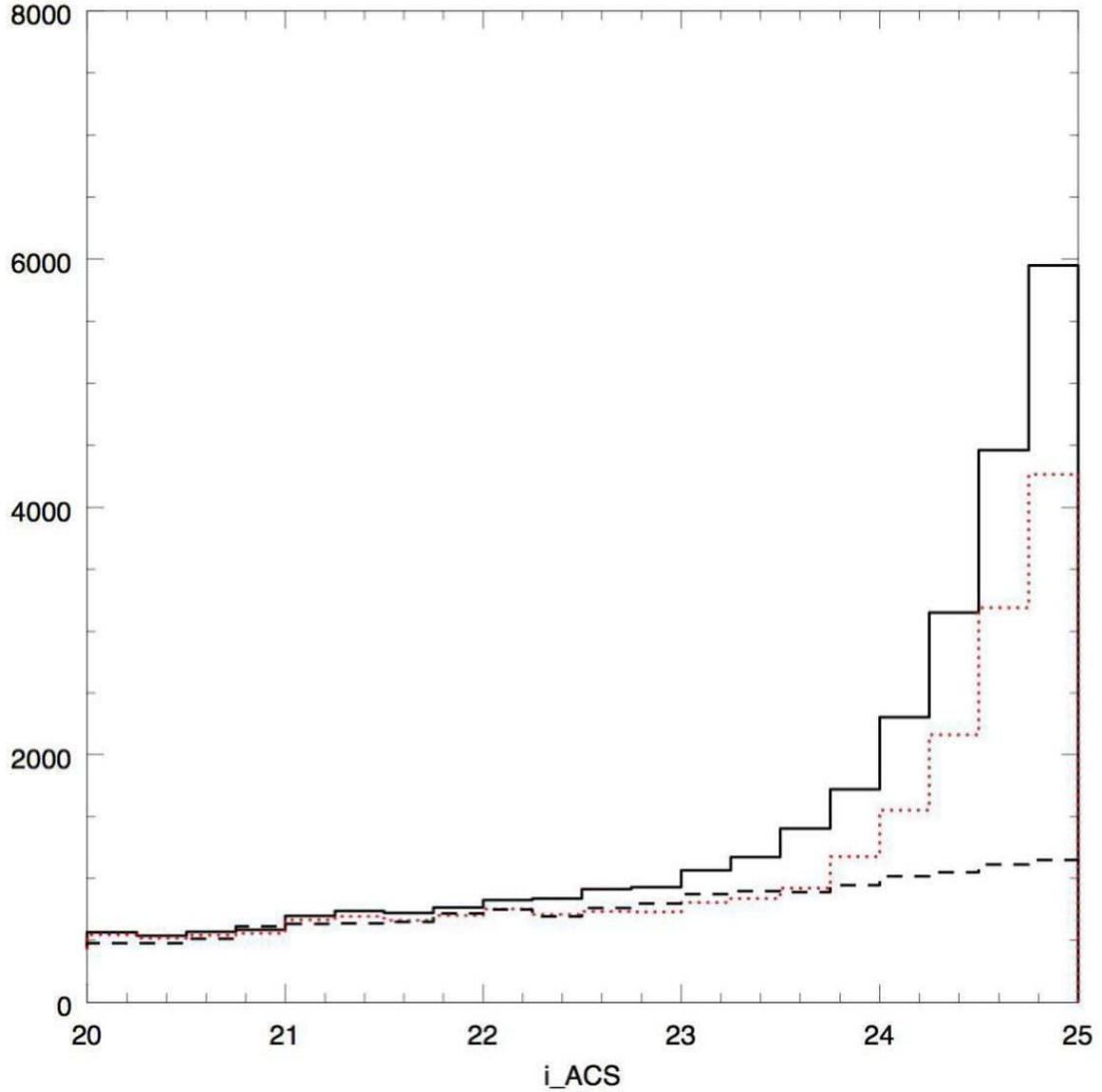}
\caption{ Star count histograms for different star selections. Solid line: objects classified as stars with the relaxed selection from the SED (see text). Dashed line: objects classified
as stars from the strict SED selection. Dotted line: star counts with the relaxed SED selection and the $i-k$ selection.}
\end{figure}



\begin{figure}
\label{cmd-data-megacam}
\plotone{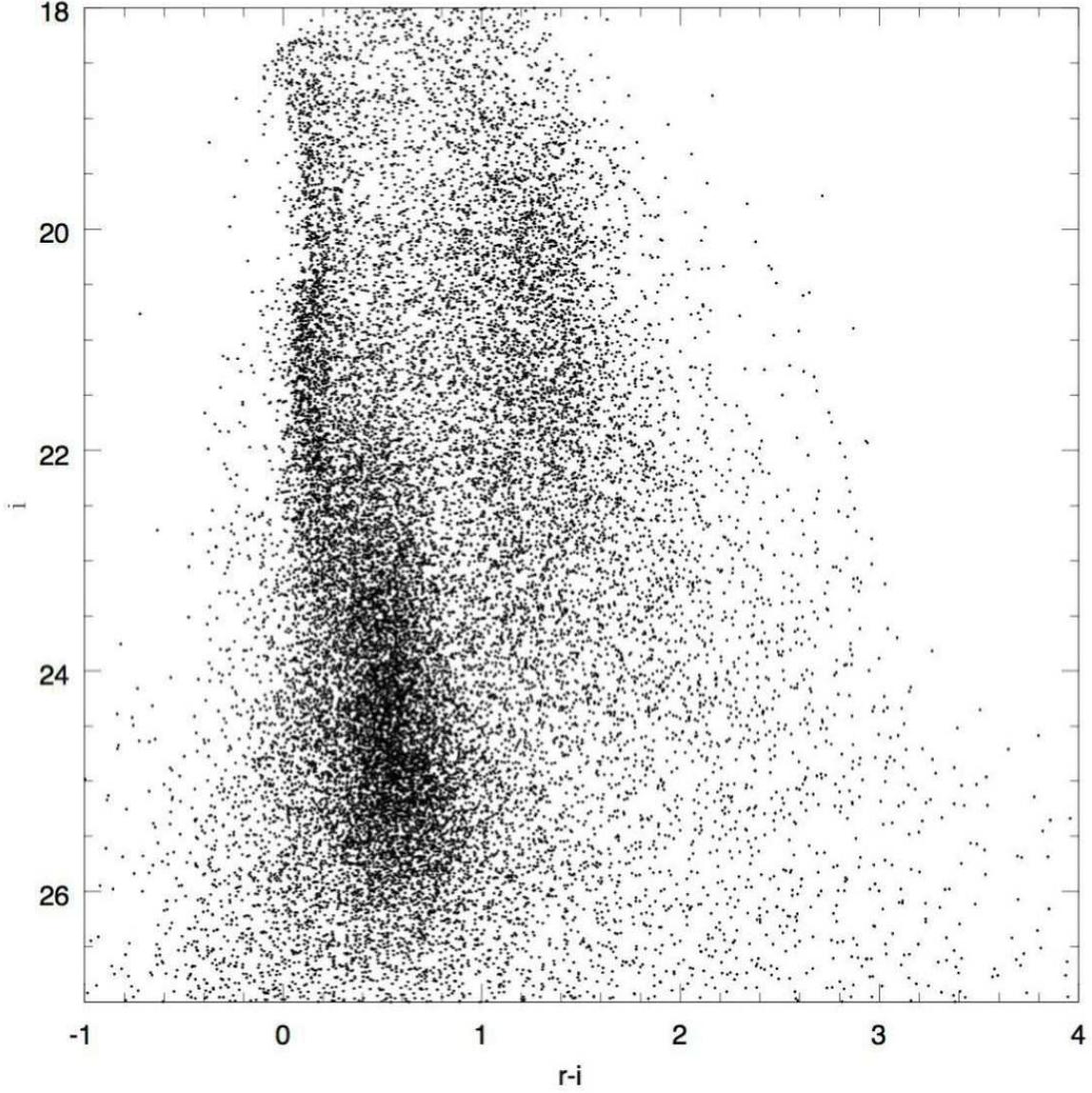}
\caption{Color-magnitude diagram $i/r-i$ (in the Megacam system, as explained in text) of observed stars with the SED-based selection criterion applied. }
\end{figure}
\clearpage

\begin{figure}
\label{fig-cmd-model-std-27}
\plotone{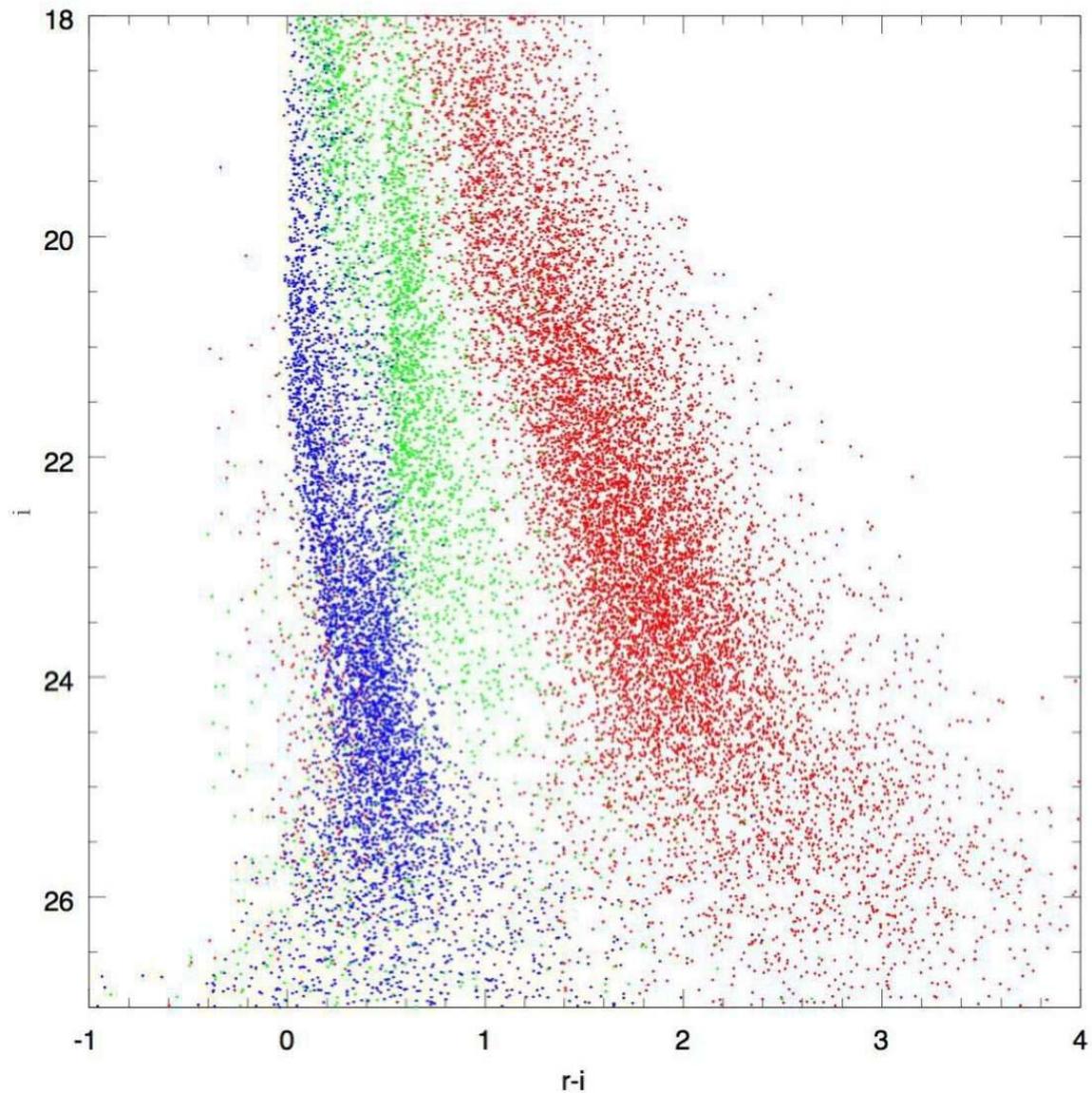}
\caption{Color-magnitude diagram $i/r-i$ simulated by the standard Besancon Galaxy model in the Cosmos field. Different stellar populations are coded in color: red: thin disk, blue: spheroid, green: thick disk.   Notice that the model is successful in reproducing the rise in counts at $I>24$, but not the
detailed color distribution.  Nonetheless, we believe that this population does not arise from
a streamer or dwarf spheroidal along the line of sight.
}
\end{figure}
\clearpage

\begin{figure}
\label{fig-hist-comp}
\plotone{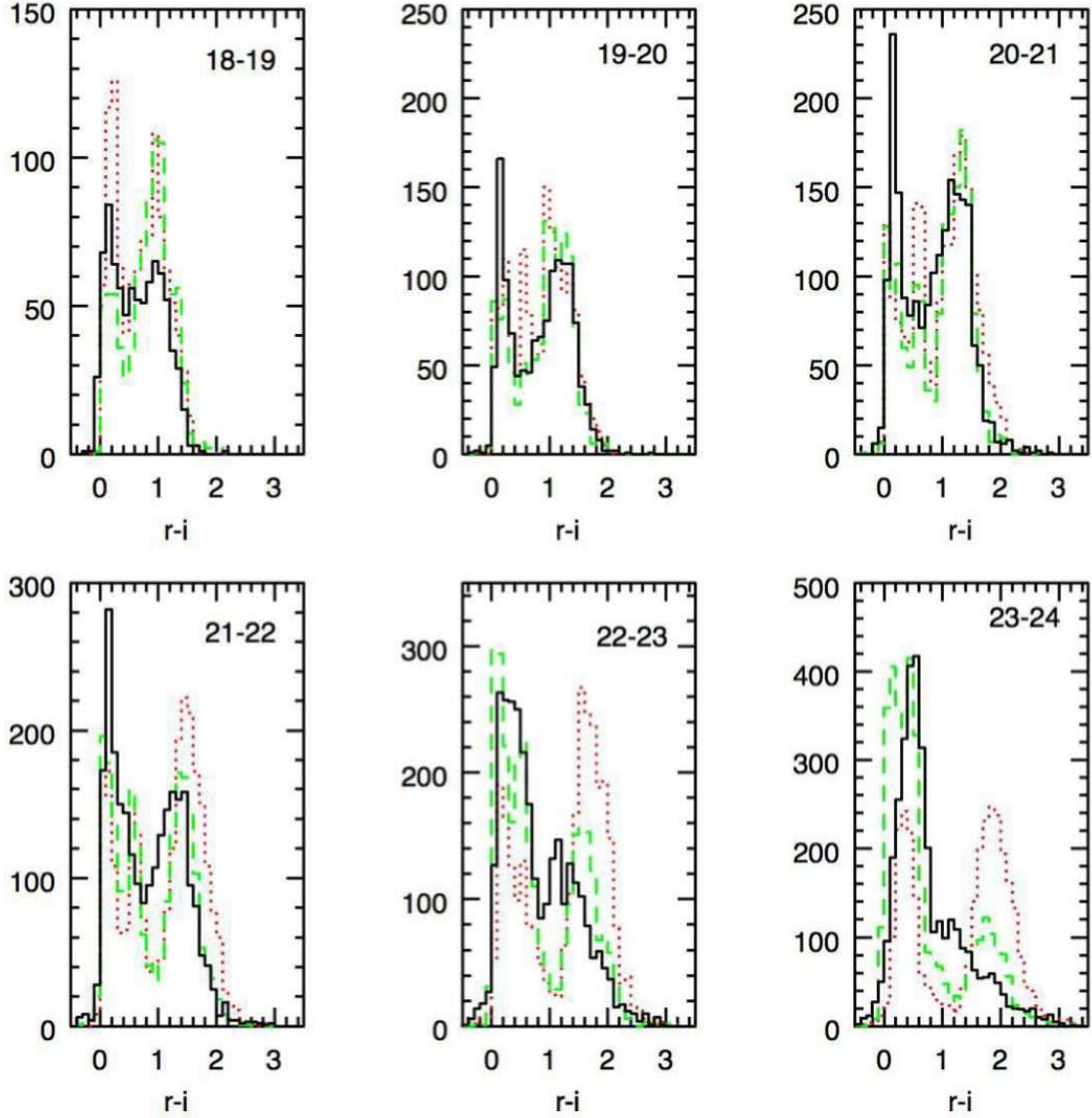}
\caption{Distribution in color $r-i$ of observed and simulated stars in different magnitude intervals. Solid  line: data. Red dotted line: standard Besan\c{c}on Galaxy model. Green dashed line : alternate model as presented in the text. }
\end{figure}

\begin{figure}
\label{fig-hist-pop}
\plotone{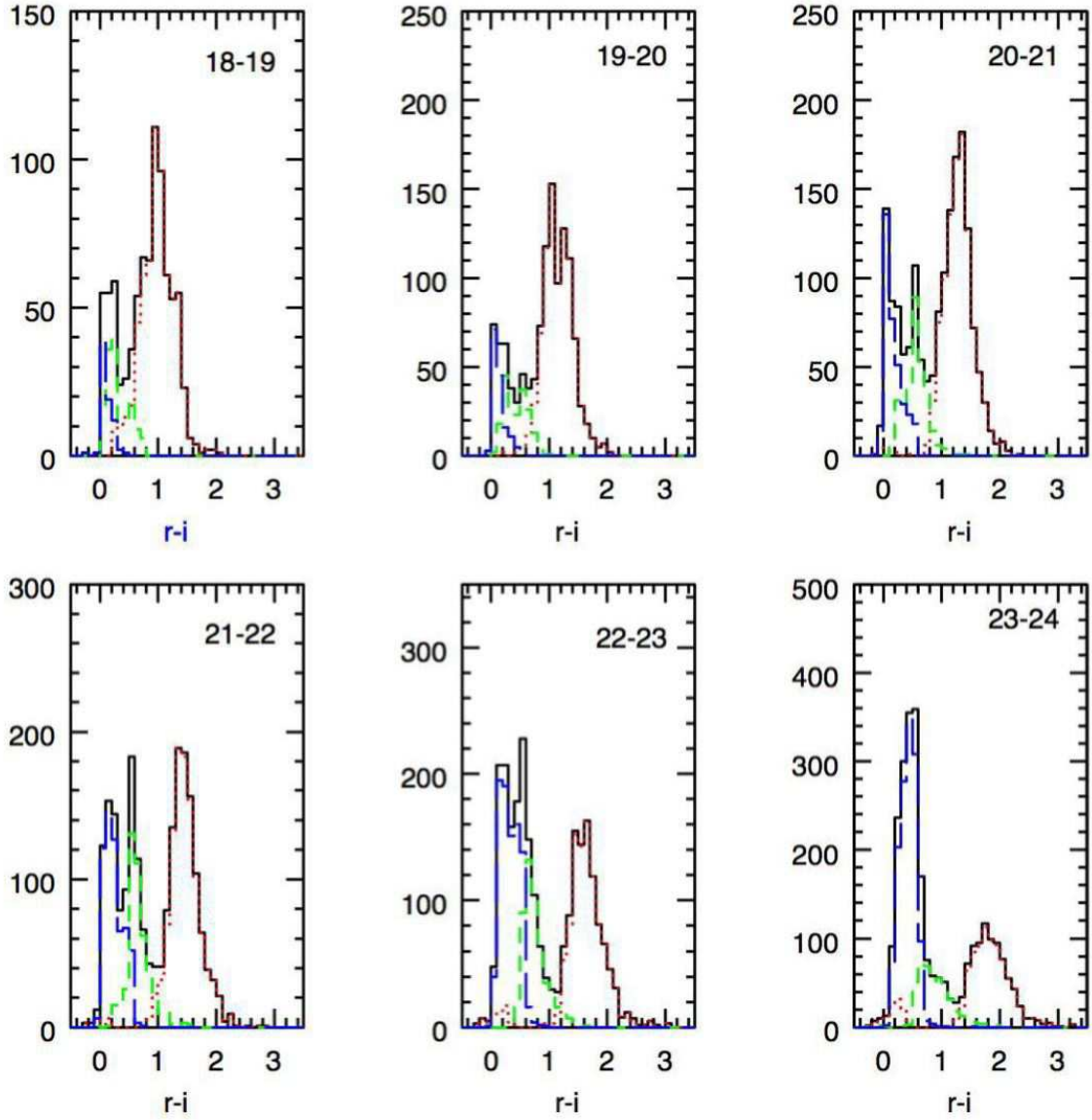}
\caption{Distribution in color $r-i$ of simulated stars from the alternate model (solid line) in different magnitude intervals. Populations are identified by colors:  thin disk in the right peak (red dotted line), thick disk in the middle (green short dashed lines), spheroid in the blue peak (blue long dashed lines). }
\end{figure}

\begin{figure}
\label{halo-density}
\plotone{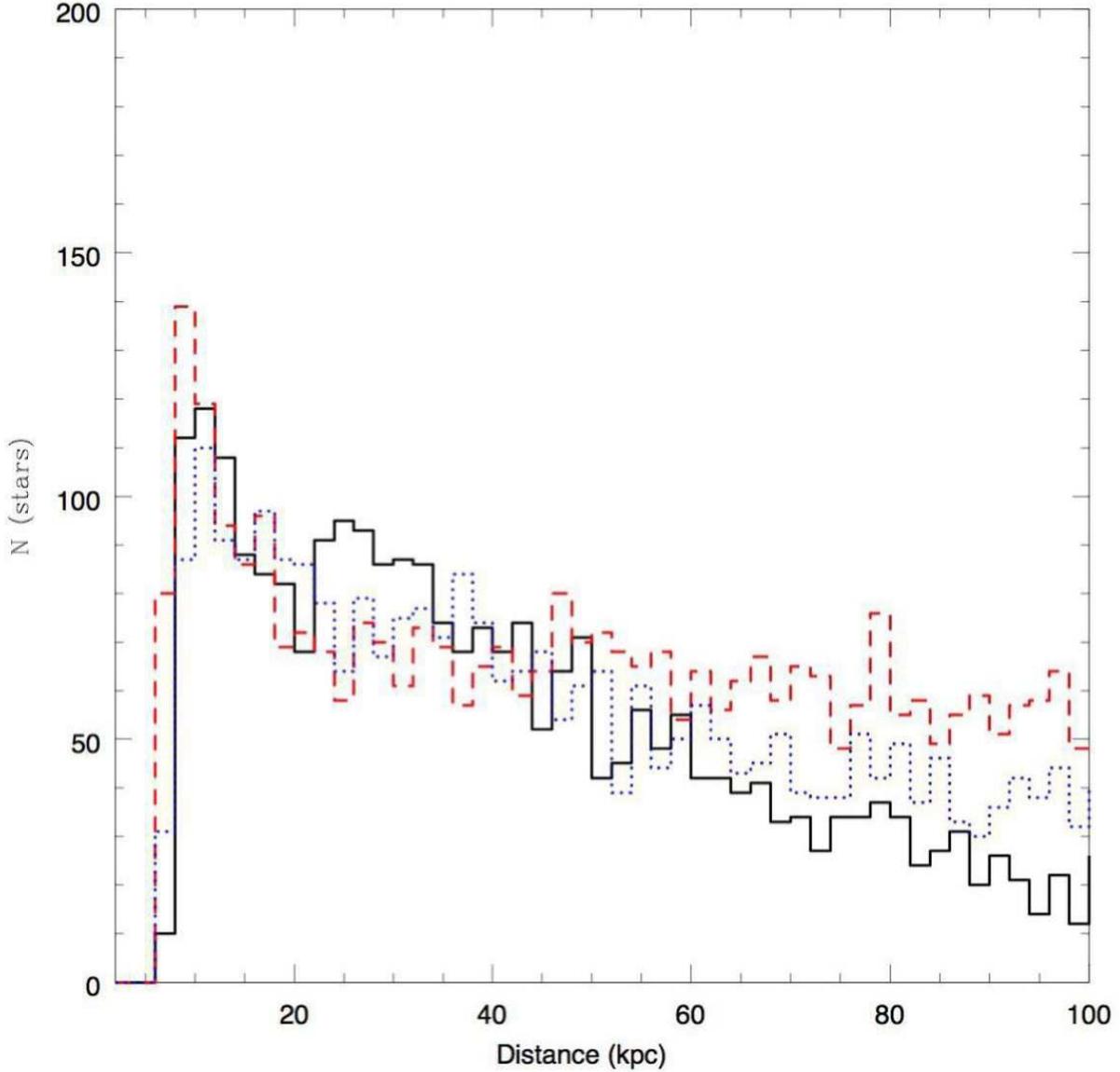}
\caption{Distribution of the distance estimator (see text) in data (solid line) and model simulations (red dashed line for the standard model with a power law exponent of 2.44, and blue dotted line for a power law exponent of 3). Distances less than 16 kpc are dominated by the thick disk turnoff rather than the halo one.  The exponent of 3. seems to better fit the data at least to distance of 70 kpc. Further than this
it is still an overestimate which might indicates the detection of a smooth edge of the spheroid.}
\end{figure}
\clearpage

\begin{figure}
\label{fig-surdens}
\plotone{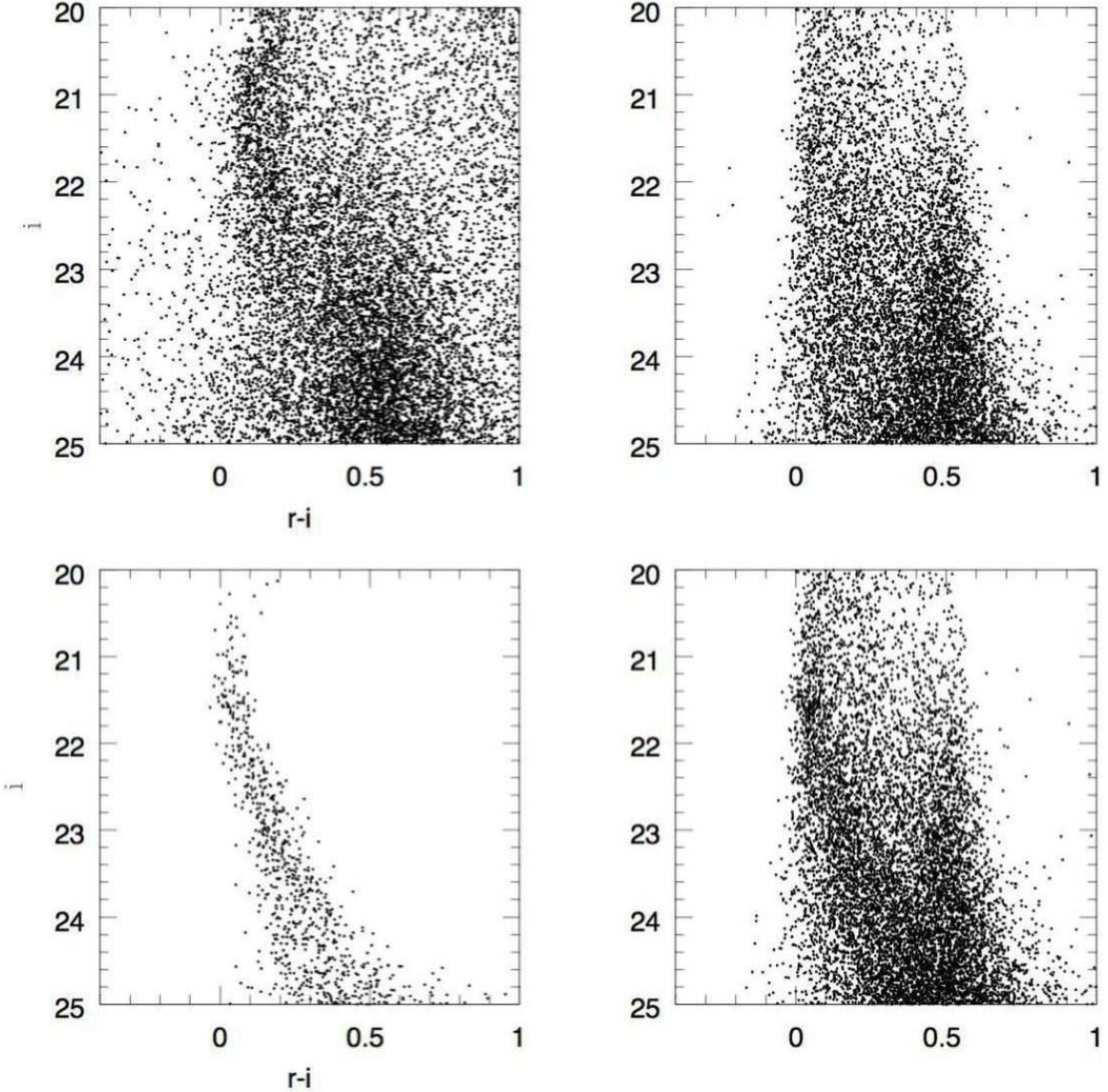}
\caption{Color-magnitude diagrams for the spheroid. Upper left: Cosmos data. Upper right: Model simulation
without an added over density for stars of the spheroid with a power law exponent of 3. Lower left: simulation of a
population similar to the spheroid (age and metallicity) at distance between 22 and 34 kpc with
same density normalisation as the spheroid. Lower right: superposition of the simulated spheroid with the overdensity. The spheroid + added turnoff model does match the data near the turnoff region but appears to overestimate the number of main sequence objects at $i>$23
and $r-i >$0.3.  The gap at $i \sim 23$ and $r-i \sim 0.25$ would be unphysical for 
a monotonically rising mass function.}
\end{figure}

\begin{figure}
\label{fig-cmd-model-alt2-27}
\plotone{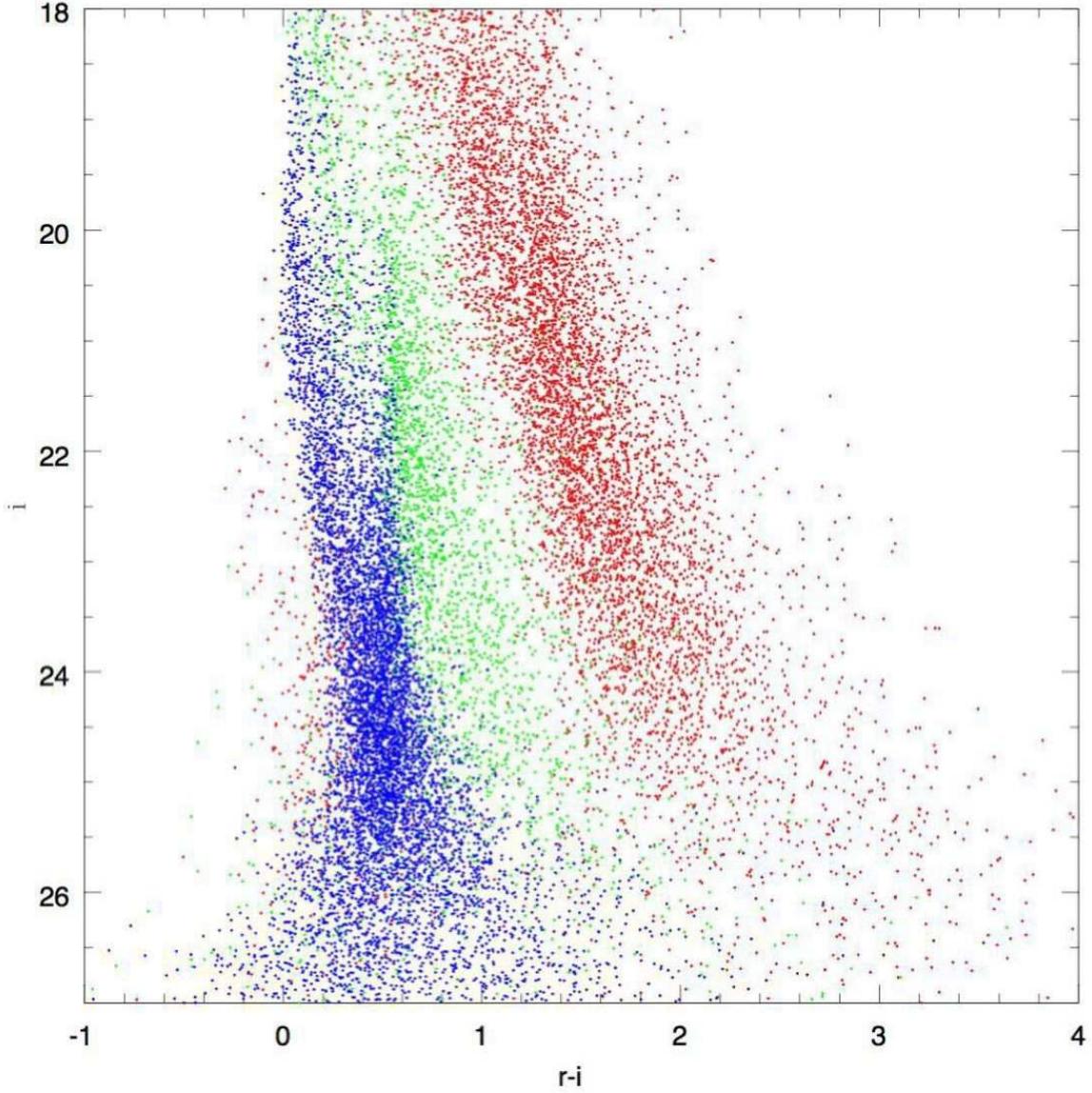}
\caption{Same as figure~5 but applying an alternative
choice for the IMF slopes for the
thin disk, the thick disk and with a scale height of 1200 pc for the thick disk (see text).}
\end{figure}

\begin{figure}
\label{fig_NG_model}
\plotone{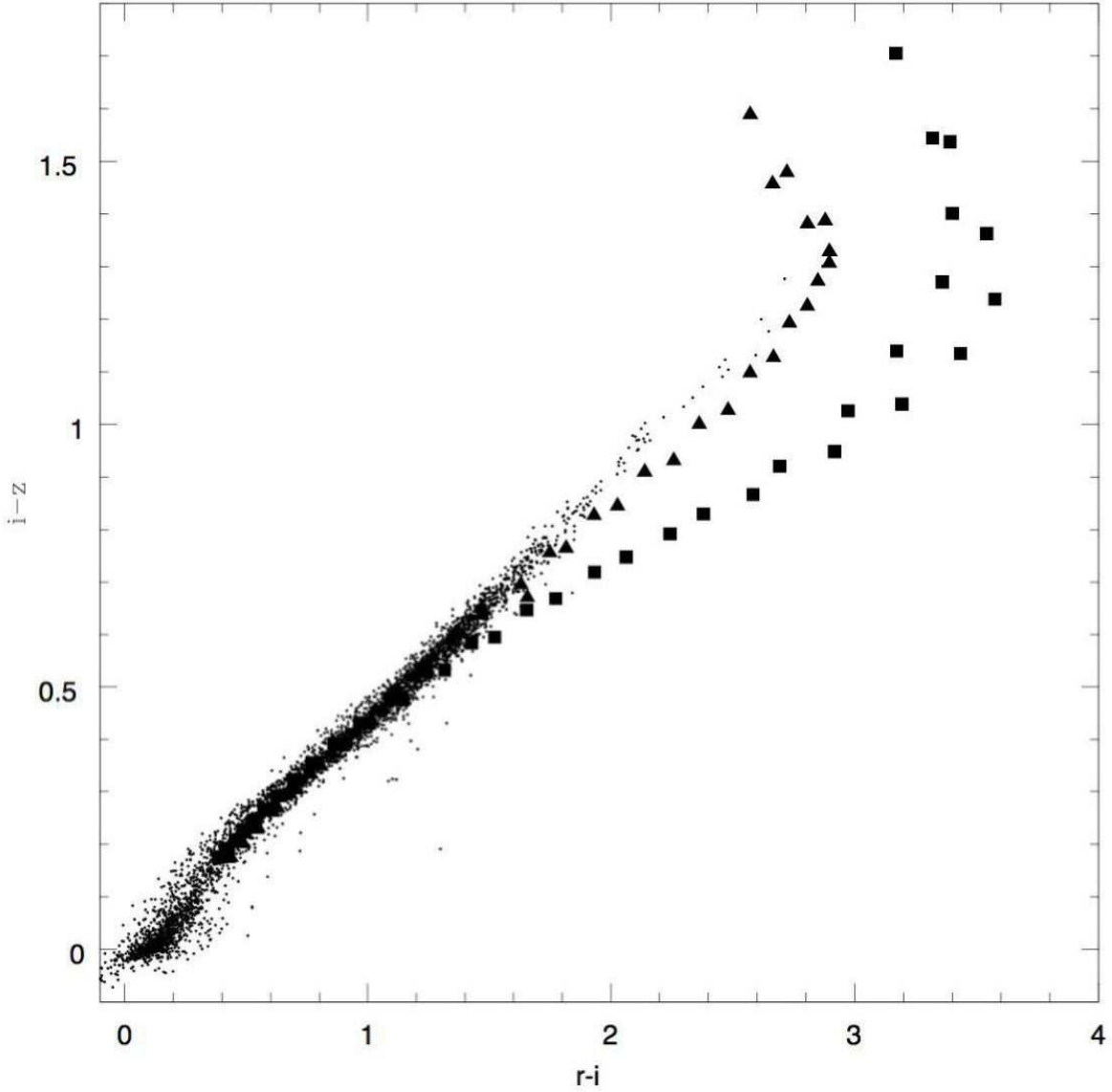}
\caption{Color-color diagram of the sample with $i<24$ in the CFHTLS photometric system. Dots: COSMOS sample. Triangles : NextGen models of $T_{\mbox{eff}}<4400 K$ and log g$>$3.5 at solar metallicity. Square: Same at metallicity [Fe/H]=-1.0. These atmosphere models deviate from the data at $T_{\mbox{eff}}<3000 K$. Models seems to be too much sensitive to metallicity since the data sequence is so narrow. Moreover if data have in the mean a solar metallicity (as expected) then the models are systematically too red by about 0.2 magnitude at $r-i>2$.
 }
\end{figure}
\clearpage

\begin{figure}
\label{white-dwarf-megacam}
\plotone{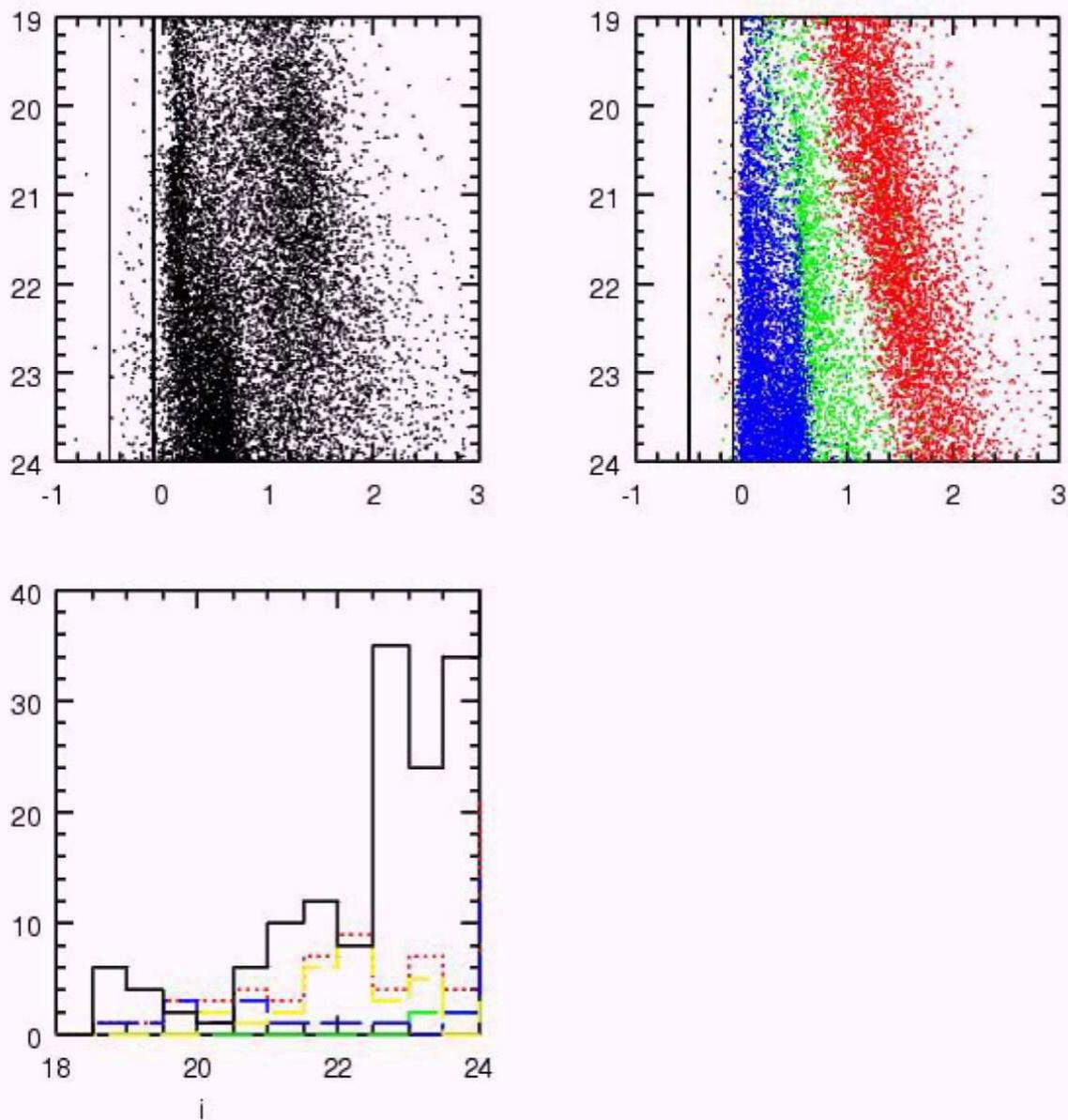}
\caption{White dwarf selection and density. Upper panel, left: data color-magnitude diagram. The white dwarf sequence selection is indicated by 2 vertical lines  $-0.5<r-i<-0.08$ placed in order to avoid a contamination by the spheroid turnoff. Upper panel, right: simulated color-magnitude diagram with color coded populations : thin disk in red, thick disk in green, spheroid in blue adn the same vertical lines for white dwarf selction. Lower panel: histogram of the number of selected objects as a function of magnitude. Solid line: Data, dotted red line: model simulation for all populations, green short dashed line: thick disk white dwarfs, blue long dashed line: spheroid white dwarf population, yellow dashed line: thin disk white dwarfs. This plot shows that the thin disk population of white dwarfs dominates at all magnitudes. The peak in the data at $i<19$ is an artefact due to saturation. In the data it seems that a contamination by QSOs affects the counts at $i>22.5$}
\end{figure}
\clearpage

\begin{figure}
\label{tdwarf}
\plotone{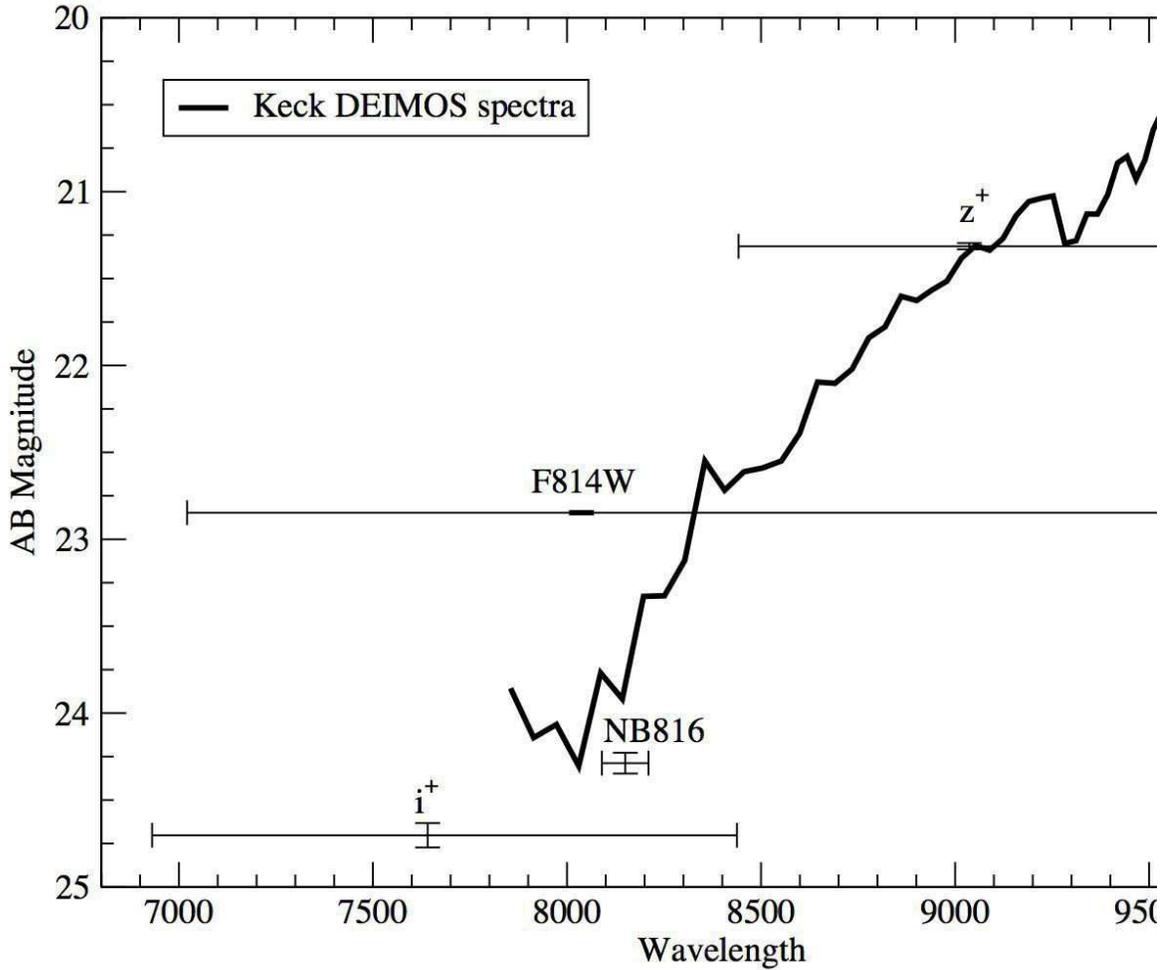}
\caption{Spectrum of a T dwarf obtained using Keck/Deimos (3x5 min exposure).  The T dwarf identification is
based on the water absorption at 9200\AA; it is very clearly seen in the unsmoothed spectra.  The star was undetected in $u$ through$ r$ and was selected as a candidate luminous high redshift object based on its red colors.  Future proper motion data  will find large numbers of these cool stars in the {\sl COSMOS} field.
}
\end{figure}


\begin{thebibliography}{}



\bibitem[Allard et al.(1997)]{Allard97} Allard, F., Hauschildt, P.~H., Alexander, D.~R., \& Starrfield, S.\ 1997, \araa, 35, 137
\bibitem[Arnouts et al.(1999)]{arnouts99} Arnouts, S., Cristiani, 
S., Moscardini, L., Matarrese, S., Lucchin, F., Fontana, A., \& Giallongo, 
E.\ 1999, \mnras, 310, 540 
\bibitem[Aussel et al.(2007)]{Aussel07} Aussel, H. et al.\ 2007, \apjs, in press
\bibitem[Belokurov et al.(2006)]{bel-fld06} Belokurov, V., et al.\ 2006, \apjl, 642, L137 
 \bibitem[Bertin \& Arnouts(1996)]{Bertin}Bertin E., Arnouts S., 1996, A\&A 117, 393 
 \bibitem[Bergbush \& VandenBerg(1992)]{Bergbush} Bergbush, P. A., \& VandenBerg, D. A. 1992, \apjs, 81, 163
 \bibitem[Bruzual \& Charlot(2003)]{bc03} Bruzual, G., \& Charlot, S.\ 2003, \mnras, 344, 1000 
\bibitem[Bullock \& Johnston(2005)]{Bullock2005} Bullock, J.S., Johnston, K.V., 2005, ApJ 635 931    
\bibitem[Capak et al.(2007)]{Capak} Capak, P. et al. 2007, \apjs,  in press
\bibitem[Cristiani et al.(2004)]{Cristiani} Cristiani, S., et  al.\ 2004, \apjl, 600, L119 
\bibitem[Coleman et al.(1980)]{cww80} Coleman, G.~D., Wu, C.-C., \& Weedman, D.~W.\ 1980, \apjs, 43, 393 
\bibitem[Dohm-Palmer et al.(2001)]{dohm01} Dohm-Palmer, R.~C., et al.\ 2001, \apjl, 555, L37 
\bibitem[Fellhauer et al.(2006)]{fel06} Fellhauer, M., et al.\ 2006, \apj, 651, 167 
\bibitem[Ferguson et al.(2002)]{ferg02} Ferguson, A.~M.~N., 
Irwin, M.~J., Ibata, R.~A., Lewis, G.~F., \& Tanvir, N.~R.\ 2002, \aj, 124, 1452
\bibitem[Zheng et al.(2004)]{Gould} Zheng, Z., Flynn, C., Gould, A., Bahcall, J.~N., Salim, S.\ 2004, \apj, 601, 500 
\bibitem[Harris et al.(1997)]{harris97} Harris, W.~E., et al.\  1997, \aj, 114, 1030 
\bibitem[Ilbert et al.(2006)]{ilbert06} Ilbert, O., et al.\ 
2006, \aap, 457, 841 
\bibitem[Ivezic et al.(2004)]{ivezic2004} Ivezic, Z., Lupton, R., 
Schlegel, D., Johnston, D., Gunn, J., Knapp, J., Strauss, M., \& Rockosi, 
C.\ 2004, ASP Conf.~Ser.~317: Milky Way Surveys: The Structure and 
Evolution of our Galaxy, 317, 179 
\bibitem[Larson(2005)]{Larson}  Larson, R.~B., 2005, MNRAS 359, 211
\bibitem[Leauthaud et al.(2007)]{Leauthaud} Leauthaud A. et al.\ 2007, \apjs, in press
\bibitem[Lejeune et al.(1997)]{Lejeune97} Lejeune Th., Cuisinier F., Buser R., 1997, AAS 125, 229
\bibitem[Mobasher et al.(2007)]{Mobasher} {Mobasher}, B. et al.\ 2007, \apjs, in press
\bibitem[Pickles(1998)]{pic98} Pickles, A.~J.\ 1998, \pasp, 
110, 863 
\bibitem[Reyl\'e \& Robin(2001)]{Reyle2001} Reyl\'e, C. Robin, A.C., 2001, Astron. Astrophys., 373:886
\bibitem[Rhodes et al.(2006)]{rho06} Rhodes, J. R. \etal~ 2007, \apjs, in press     
\bibitem[Robin et al.(2000)]{Robin2000} Robin A.~C., Reyl{\' e} C., \& Cr{\' e}z{\' e} M.\ 200
0, A\&A, 359, 103
\bibitem[Robin et al.(2003)]{Robin2003} Robin A.~C., Reyl\'e C., Derri\`ere S., Picaud S., 200
3, A\&A 409, 523
\bibitem[Robin et al.(2004)]{Robin2004} Robin, A.C., Reyl\'e, C., Derriere, S., Picaud, S., 2004, Astron. Astrophys. 416, 157.
\bibitem[Sanders et al.(2007)]{sanders07} Sanders, D.~B. et al.\ 2007, \apjs, in press
\bibitem[Schlegel et al.(1998)]{Schlegel} Schlegel, D.J., Finkbeiner, D.P., \& Davis, M. 1998, ApJ 500, 525
\bibitem[Scoville et al.(2007a)]{sco07a}Soville, N.~Z., Abrhama, R.~G., Aussel, H. et al. 2007a, \apjs, in press
\bibitem[Scoville et al.(2007b)]{sco07b}Scoville, N.~Z., Aussel,H., Brusa, M., Capak, P. et al. 2007b, \apjs, in press
\bibitem[Schultheis et al.(2006)]{Schultheis2006} Schultheis, M., Robin A.~C., Reyl\'e C., H.J. McCracken, E. Bertin, Y. Mellier, and 0. Le Fvre. 2006, A\&A 447, 185-198
\bibitem[Vivas \& Zinn(2006)]{Vivas} Vivas, A.K., \& Zinn, R., 2006, AJ 132, 714


\end{thebibliography}
 \end{document}